# Controllable Bistability in Dual-Fiber Optical Trap in Air

Aoxi Chen [a,b,1], Xinlin Chen [a,b,1], Siyuan Rao [a,b], Hui An [a,b], Yingying Song [a,b], Tengfang Kuang [a,b], Wei Xiong [a,b], Xiang Han [a,b], Zhongqi Tan [a,b], Guangzong Xiao [a,b,*], Hui Luo [a,b]

[a] *College of Advanced Interdisciplinary Studies, National University of Defense Technology, Changsha Hunan, 410073, China*
[b] *Nanhu Laser Laboratory, National University of Defense Technology, Changsha Hunan, 410073, China*
[1] *These authors contributed equally.*
* *xiaoguangzong@nudt.edu.cn*

**Abstract**: The dual-fiber optical trap, owing to its high sensitivity and facile miniaturization, holds significant actual application value in fields such as high-precision metrology of mechanical quantities and biological manipulation. The positional stability of the trapped particle is pivotal to system performance, directly setting the measurement noise floor and operational precision. In this work, we observe bistability and hysteresis in the axial equilibrium position of a 10-μm diameter $SiO_2$ microsphere. This bistability arises from optical interference between the fiber ends and the microsphere, creating multiple potential wells. Experimental results demonstrate that the microsphere's transition rate can be effectively modulated through precise control of the trapping laser power. Furthermore, the incorporation of transverse misalignment has effectively eradicated bistability, thereby substantially improving positional stability throughout the entire optical trapping region. This suppression successfully reduced the system's residual positional uncertainty to the thermal noise limit. Consequently, this research will enhance the precision of microparticle manipulation and the sensitivity of sensing in dual-fiber optical trap systems.

## 1. Introduction

Optical tweezers have revolutionized micro-manipulation by enabling non-contact trapping and control of particles[1–4]. This powerful technique finds broad applications in life sciences[5–7], quantum physics[8–10], precision measurement[11–13] and nanoscale material characterization[14,15]. Among its various implementations, the dual-fiber optical trap (DFOT) stands out for its compact structure and enhanced stability, showing significant potential for applications in micro-force sensing[16,17], accelerometry[18,19], biological[20–23] and physical[24–26] research. Chen et al. revealed motion transitions such as spiral and orbital rotation under beam offset, findings that aid micro-force sensing[27]. In the context of accelerometry, Han et al. developed a feedback-controlled levitated optomechanical accelerometer using a dual-beam trap, demonstrating high sensitivity and stability in practical environments[28]. Furthermore, B. J.

Black et al. demonstrated the rotation of human smooth muscle cells using a DFOT with intentional lateral misalignment[29]. However, the high precision required for these sensitive applications is often compromised by an intrinsic instability known as optical bistability.

Optical bistability refers to the phenomenon where a system exhibits two distinct equilibrium states under identical external conditions[30–33]. In multi-particle systems, optical binding can induce such bistable states[34,35]. This behavior is well-documented in optically bound clusters or linear chains, where inter-particle coupling creates complex potential energy landscapes. However, in sensing applications that typically involve a single trapped microsphere, the origin and dynamics of bistability remain poorly understood. This phenomenon has been advantageously exploited in applications such as information storage[36–38], photonic devices[39–41], and energy conversion[42–44]. However, its unintended emergence in single-particle DFOT systems is problematic, causing stochastic positional jumps that severely undermine measurement precision. Recent studies on systems like intracavity dual-beam optical tweezers have shown that self-feedback mechanisms can create bistable potential wells, and that adjusting parameters (e.g., foci offset) can suppress these destabilizing effects[45]. Moreover, Aaron Schäpers et al. manipulated the polarization state of coherent counter-propagating beams in a DFOT to control interference fringes, thereby achieving highly efficient trapping of microparticles in air[46]. Despite these insights, systematic research on the bistable behavior of a single microsphere in a non-interferometric DFOT remains lacking.

This study demonstrates controlled manipulation of optical bistability within a non-interferometric DFOT system. Intrinsic bistable behavior was observed in a $SiO_2$ microsphere trapped in ambient air using this platform. A systematic investigation was carried out to explore the formation mechanism of bistability and its associated characteristic features. By precisely regulating the trapping laser power at both fiber terminals, the transition rate was reduced to achieve stable trapping. Furthermore, introducing transverse misalignment in the dual-fiber optical trap effectively suppressed bistability. Together, these methods significantly enhanced the positional stability of the trapped microspheres in the DFOT system. Consequently, the approach provides a robust foundation for high-precision optical manipulation in both air and vacuum environments. It offers broad implications for advancing optical trapping-based sensing, precision metrology, and the development of advanced integrated photonic systems requiring precision particle control.

## 2. Experimental setup and Principle

*2.1 Experimental setup*

The experimental configuration incorporated a 980 nm laser source whose output was first coupled into a single-mode fiber (Corning Hi-1060). The beam was subsequently divided into two equal-intensity arms via a 50:50 fiber splitter to implement DFOT. Prior to injection into opposing ports of the DFOT system, each beam underwent independent power tuning through variable optical attenuators (VOAs). By deriving both trapping beams from a common laser source through beam-splitting architecture, inherent power differential stability was guaranteed. This counter-propagating beam geometry established a robust equilibrium in the central optical potential well[47], enabling three-dimensional levitation of microparticles in air as demonstrated in Fig. 1a.

The single-mode fibers were coated with a $Ta_2O_5$ anti-reflection layer to minimize the impact of back-reflected light, achieving a reflectance of 0.17%. The fiber-to-fiber spacing was maintained at 150 μm to ensure stable trapping at the center of the optical potential well. Additionally, the optical path difference between the two beams was intentionally designed to exceed the laser's coherence length significantly, thereby effectively suppressing inter-beam interference. A $SiO_2$ microsphere with a diameter of 10.02 μm was stably levitated within the trap. SEM characterization verified that the microsphere exhibited smooth surfaces without contaminants (Fig. 1c). The microsphere's position was monitored in real-time using the centroid method, whereby its displacement was calculated by tracking the center of mass of the image captured by a CCD camera. Consequently, the interplay of reflections from the fiber end-faces and the microsphere surface created a complex optical potential landscape (Fig. 1b). Notably, any displacement of the microsphere within the trap immediately altered the optical cavity lengths of both resonators; for instance, a movement to the right increased the length of cavity 1 while decreasing that of cavity 2. These changes modified the intracavity optical power distribution, which in turn adjusted the optical force distribution to form the observed potential wells.

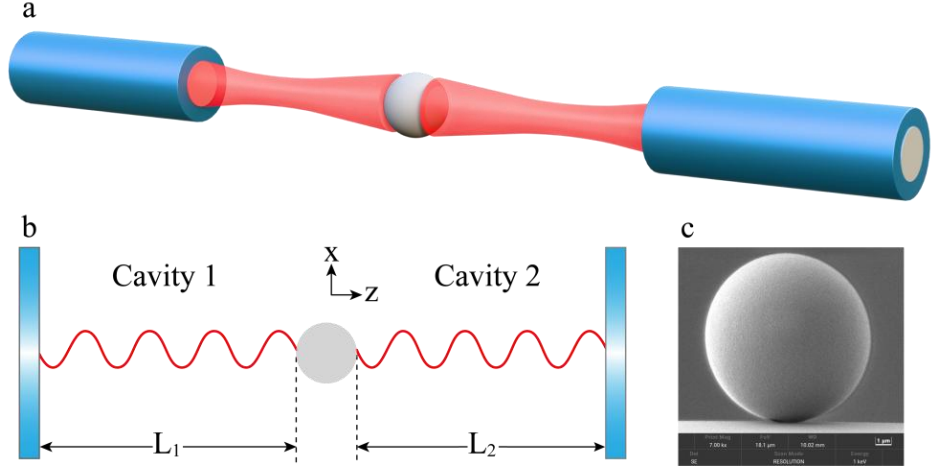

Fig. 1. **a.** Model of the DFOT system. **b.** Schematic diagram of the dual-cavity model in the DFOT system. **c.** SEM image of the surface morphology of a SiO$_2$ microsphere.

*2.2 Principle*

Given that the radius of the levitated microsphere is significantly smaller than the curvature radius of the fiber end face, both optical cavities can be approximated as resonators formed between planar and spherical surfaces. Their finesse $F_n$ can be expressed as:

$$F_n = \frac{\pi \sqrt[4]{R_0 R_n}}{1 - \sqrt{R_0 R_n}}, (n = 1, 2) \tag{1}$$

where $R_0$ and $R_n$ represent the reflectivity of the microsphere and the two fiber end faces, respectively. The intracavity optical power $P_n$ (n=1,2) of the two cavities follows a modulated relationship:

$$P_n = \frac{1}{1 - R_0} \cdot \frac{P_0}{1 + (\frac{2F_n}{\pi})^2 \sin^2 \left[ \frac{2\pi L_n}{\lambda} + \varphi_n \right]}, (n = 1, 2) \tag{2}$$

where $P_0$ represents the output power of the laser, $L$ is the effective cavity length, and the reflection phase shift $\varphi_n$ is derived from the Fresnel equations:

$$\varphi_n = \arctan(\frac{n_1 \cos\alpha_{in} - n_2 \cos\alpha_{tn}}{n_1 \cos\alpha_{in} + n_2 \cos\alpha_{tn}}), (n = 1, 2) \tag{3}$$

where $n_1$ and $n_2$ are the refractive indices of the microsphere and the ambient medium, respectively, while $\alpha_{in}$ and $\alpha_{tn}$ denote the incidence and refraction angles, respectively, of the single laser beam incident upon the microsphere surface within two cavities.

The effective number of reflections within the cavity can be calculated using Eq. (1). Due to the relatively low quality of the cavity, the reflection count remains limited, allowing us to neglect the influence of beam radius variations on the power distribution. The microsphere diameter in the experiment significantly exceeds the laser wavelength $\lambda$, satisfying the ray optics approximation. Therefore, the scattering force $F_s$ and gradient force $F_g$ acting on a microsphere can be derived from key parameters such as intracavity power, operating wavelength, beam waist radius, fiber separation distance, azimuth angle, and curvature radius, following the methodologies established in ref.[48]. Subsequently, by substituting the intracavity power of cavity 1 and cavity 2 into the expressions for $F_s$ and $F_g$, we can calculate the optical force components exerted by each cavity. Finally, the total optical trapping force on the microsphere is obtained through vector summation of these components.

*2.3 Simulation*

The numerical simulations are conducted using the following parameters: Gaussian beam wavelength $\lambda_0 = 980$ nm, optical power $P_1 = P_2 = 150$ mW, beam waist radius $\omega_0 = 3$ μm, and fiber separation distance $S = 150$ μm. The system configuration included air (refractive index $n_1 = 1.0$) as the medium and a $SiO_2$ microsphere with refractive index $n_2 = 1.45$, diameter $d = 10$ μm, and surface reflectivity $R_0 = 4\%$. The fiber end-faces are coated with anti-reflection films, exhibiting a reflectivity of $R_1 = R_2 = 0.17\%$. Fig. 2a presents the simulation result of the optical field distribution within the DFOT. It clearly reveals an interference pattern resulting from the repeated reflections of the trapping beam between the fiber end-face and the microsphere.

Fig. 2b illustrates the axial optical force distribution acting on the microsphere within the range of -50 μm to 50 μm. The experimental results demonstrate that the intra-cavity interference effect induces a periodic variation in the optical force as a function of the microsphere's displacement, with a period of 490 nm, as shown in Fig. 2c. Through spatial integration of the optical force distribution, the potential energy profile in the central region of the optical trap was derived, as depicted in Fig. 2c. Analysis of the potential energy curve reveals two distinct trapping regimes: In the DFOT system, Position I and Position III are identified as stable equilibrium points corresponding to the local minima of the system's potential energy profile, while Position II represents an unstable equilibrium point associated with the local maximum of the potential energy curve. When a microsphere is trapped at the optical trap center, if both potential energy differences between Position I and II ($\Delta U_{12}$) and Position III and II ($\Delta U_{23}$) are smaller than the characteristic energy of Brownian motion, thermal fluctuations will induce stochastic transitions between Position I and Position III. This energy-competitive process ultimately manifests as a bistable phenomenon in the microsphere's positional dynamics.

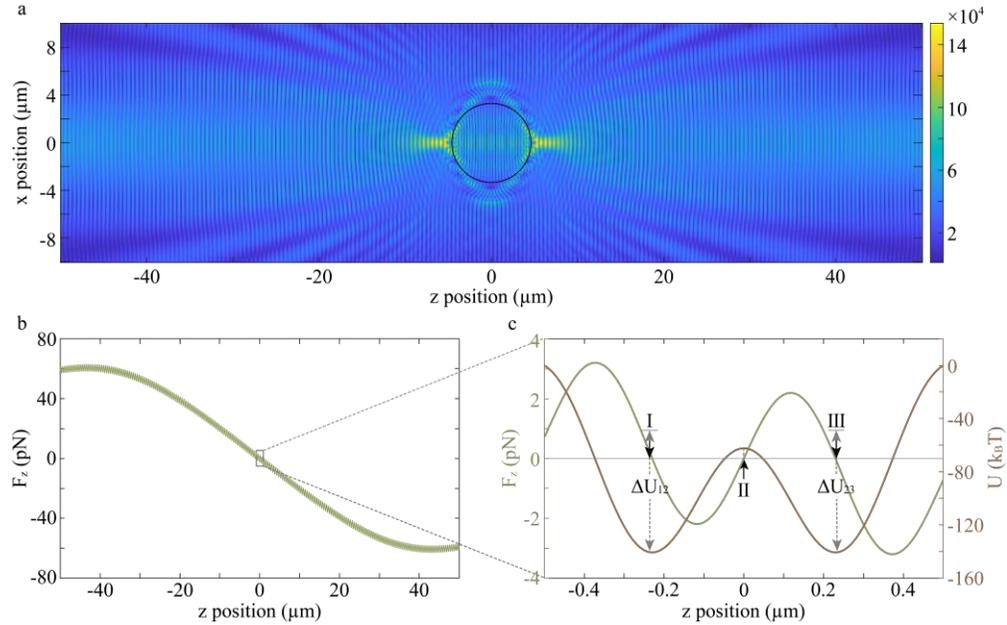

Fig. 2. Simulation Result. **a.** The optical field distribution. **b.** Distribution curve of optical force within the range of -50 μm to 50 μm. **c.** Optical force profile at the trap center and well distribution.

### 3. Experimental Results and Discussion

To validate the theoretical predictions and systematically resolve the bistability issue in the DFOT system, we conducted controlled experiments by precisely balancing the optical power at both fiber ends. We modulated the optical power difference (Δ$P$) by periodically adjusting the control voltage of the VOAs at 20 s intervals. This controlled imbalance consistently displaced the microsphere to one side. Figure 3a shows the dynamic state of the microsphere evolving with Δ$P$. As Δ$P$ varies, the microsphere is not stably trapped at the center but instead exhibits bistability, dynamically transitioning between two metastable states along the optical axis. This stochastic switching behavior between Position I and Position III is clearly illustrated in Fig. 3c.

To elucidate the mechanism underlying the switching behavior, we performed a comprehensive analysis through numerical simulations of potential energy combined with experimental investigations. Figure 3b shows the potential energy distribution and oscillation dynamics under varying Δ$P$. Under conditions of Δ$P$ = -0.8 mW, the potential energy difference at Position I exceeds the kinetic energy associated with the microsphere's Brownian motion. This confines its oscillation amplitude (green dashed lines and shaded area in Fig. 3b) to a range smaller than the distance between the stable and unstable points. Consequently, the microsphere remains stably trapped at Position I. When ΔP increases to a specific value, the system exhibits bistable behavior (Fig. 3c(i)). This originates from a pressure gradient-induced reduction in both the spatial separation and potential energy difference between Positions I and II. Once the potential energy barriers at these positions decrease below the microsphere's thermal energy, the particle exhibits dramatically enhanced oscillation amplitudes (yellow dashed lines and shaded area in Fig. 3b), facilitating its transition to Position III. Nevertheless, a restorative leftward optical force at this new position promptly returns the microsphere to its original location. This persistent bidirectional transition mechanism ultimately gives rise to the system's optical bistability.

As the $\Delta P$ further increases to 0 mW, the transition rate of the microsphere increases (Fig. 3c(ii)). A subsequent increase to 0.4 mW leads to a decrease in transition rate (Fig. 3c(iii)). Notably, within a specific range of $\Delta P$, the transition rate initially increases and then decreases. This non-monotonic behavior occurs because increasing $\Delta P$ simultaneously reduces the energy barrier between Positions I and II while raising the barrier between Positions II and III. The former enhances forward transition probability, whereas the latter suppresses reverse transitions. As the power continues to rise, the distance between the unstable point and the subsequent stable point expands, dampening the microsphere's oscillations and preventing its return (blue curve in Fig. 3b). Consequently, the microsphere becomes stably trapped at Position III (Fig. 3c(iv)). Through precise control of the $\Delta P$ between the two optical fibers, we demonstrate accurate manipulation of bistable switching and microsphere transition rates. This capability enables dynamic positioning of microspheres within the optical potential well while circumventing bistable regions, thereby enhancing positional stability.

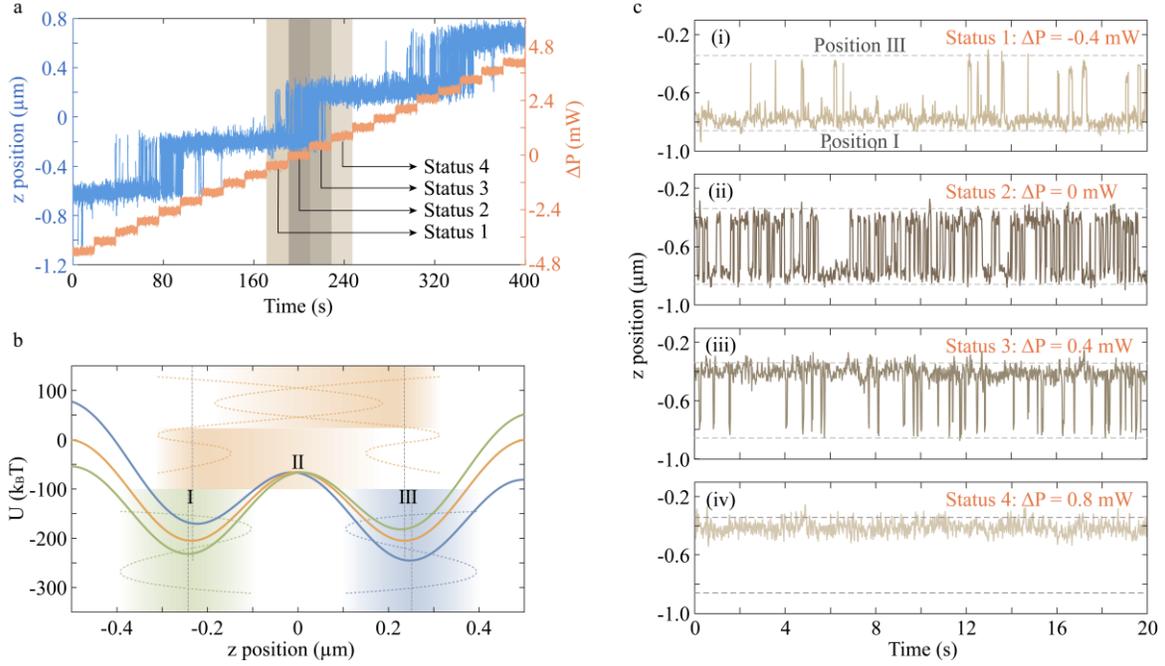

Fig. 3. Bistable Behavior of a Microsphere in the DFOT System. **a.** Experimentally measured displacement of the microsphere versus optical power. **b.** Simulated optical force distribution and vibrational states of the microsphere at the trap center under different ΔP. **c.** Motion states of microspheres under different ΔP.

The bistable behavior of the microsphere is unambiguously revealed by a characteristic hysteresis loop in the DFOT system. In our experimental protocol, ΔP is first incrementally increased and then systematically decreased. Fig. 4 presents the microsphere displacement as a function of ΔP. The green and blue data points denote the time-averaged positions (2.9 s and 3.2 s, respectively) during the ascending and descending phases of ΔP. At ΔP = -2.4 mW, the microsphere maintains stable trapping at Position I. With increasing ΔP, the system transitions from single-state trapping to bistable operation, evidenced by the microsphere's an intermediate average position between two equilibrium states. Upon further ΔP elevation, the microsphere ultimately stabilizes near Position III. The reverse process occurs during ΔP reduction, with the microsphere transitioning back to Position I. Crucially, the distinct ΔP thresholds for bistability during the ascending and descending phases give rise to the pronounced hysteresis loop in Fig. 4. This controllable hysteretic behavior holds significant promise for optical data storage and ultra-precise sensing applications. The ability to engineer the hysteresis loop's transition thresholds establishes this system as a versatile platform for developing next-generation all-optical switches with enhanced performance.

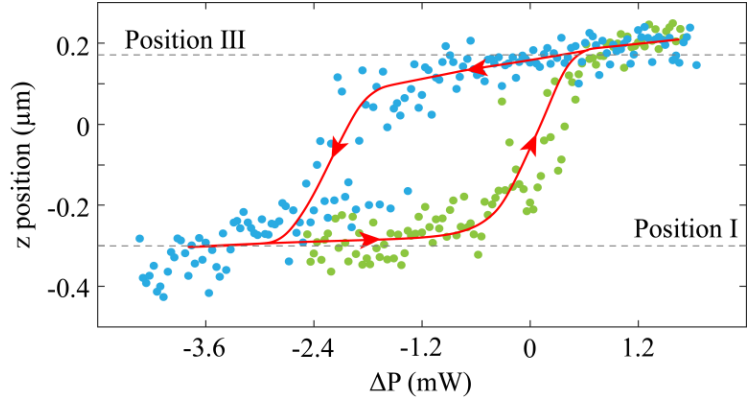

Fig. 4. Experimental observation of transition hysteresis in the DFOT System.

To address the bistability phenomenon in DFOTs and further enhance the positional stability of the microsphere within the optical trapping region, this study introduces a method based on transverse misalignment. Figure 5a illustrates a schematic of the dual-fiber trap with transverse misalignment, where the two fibers are not perfectly aligned but offset by a distance $d$ along the x-axis. The presence of transverse misalignment serves to weaken the interference effect observed between the fiber end-faces and the microsphere. However, previous studies have indicated that excessive misalignment may lead to increased jitter of the microsphere, and beyond a certain threshold, may even cause orbital rotation or escape[27]. Figure 5b compares the simulated potential energy distributions at the trap center for the cases without misalignment and with a misalignment of 3.3 μm. The results show that under misalignment, the potential well difference at the trap center is significantly reduced compared to the aligned case, indicating that introducing misalignment helps suppress bistability. As shown in Figure 5c, when $d$ = 3.3 μm, the microsphere remains stably trapped under different $\Delta P$, and no noticeable bistable transition is observed. This confirms that introducing an appropriate amount of transverse misalignment in a DFOT can effectively suppress bistability and improve the axial stability of the microsphere. Furthermore, the linear response of the microsphere displacement to power variations establishes a crucial experimental foundation for the development of high-precision optomechanical accelerometry.

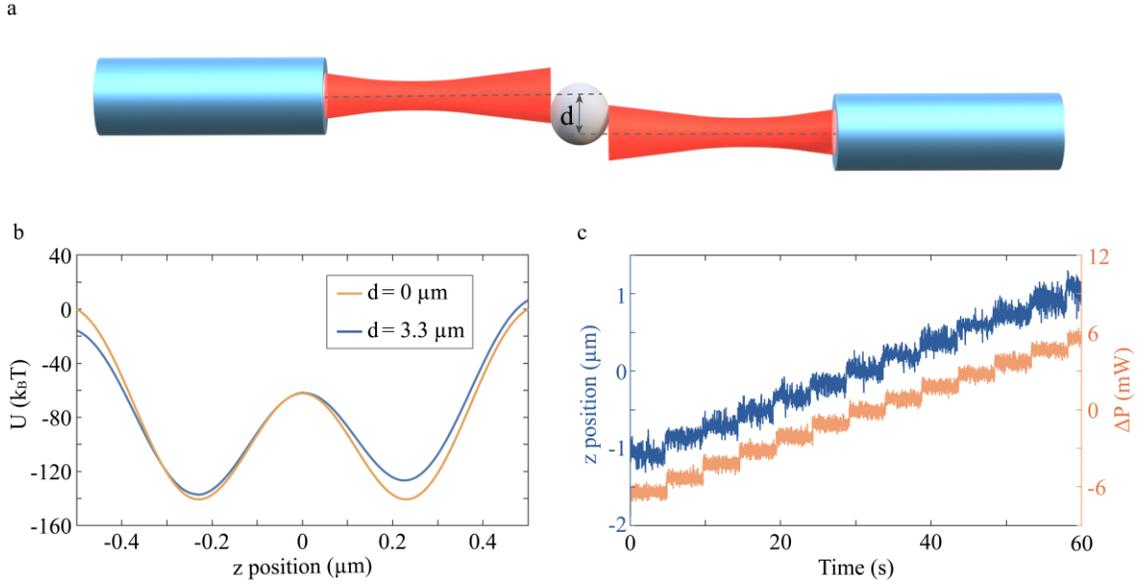

Fig. 5. Effects of transverse misalignment in a DFOT. **a.** Schematic of the misaligned trap configuration. **b.** Simulation results showing potential energy distributions for various misalignment distances. **c.** Experimental record of the microsphere displacement versus $\Delta P$ at $d$ = 3.3 μm.

## 4. Conclusions

This study investigates the bistability of a microsphere within a DFOT system, comprehensively exploring its formation mechanisms and control strategies. Through ray optics simulations, we demonstrate that this bistability stems from optical interference between the fiber end-face and the microsphere surface. This interference induces a non-uniform optical power distribution within the trap, forming a multi-well potential structure. By precisely modulating $\Delta P$ between the two fiber ends, we effectively suppress inter-well transitions of the microsphere and achieved stable trapping, alongside the observation of a distinct hysteresis effect. Furthermore, to enhance the microsphere's stability, we introduce a transverse misalignment into the DFOT. This adjustment reduces the potential energy barrier, thereby completely eliminating the bistable behavior.

Our work establishes a theoretical foundation and offers practical solutions for addressing bistability in DFOT systems. The ability to suppress bistable transitions and maintain steady trapping highlights its utility in applications requiring high spatial stability, such as high-resolution force, acceleration sensing and biological manipulation. This enhanced stability not only advances optical trapping technologies but also opens up new pathways for micro/nanoscale metrology and weak signal detection in interdisciplinary research.
**Funding**

Quantum Science and Technology-National Science and Technology Major Project (Grant No. 2024ZD0301000); National Natural Science Foundation of China (12474368,12404420); Science Fund for Distinguished Young Scholars of Hunan Province (2024JJ2055); The Key Science and Technology Breakthrough Program of Hunan Province (2023ZJ1010).

**Acknowledgments**

The authors wish to thank the anonymous reviewers for their valuable suggestions.

**Disclosures**

The authors declare no competing financial interest.

**Data availability**

Data underlying the results presented in this paper are not publicly available at this time but may be obtained from the authors upon reasonable request.